# Electroosmotic flow of viscoelastic fluids in deformable microchannels

Siddhartha Mukherjee[1], Sunando DasGupta[1,2], Suman Chakraborty[1,3]

*[1]Advanced Technology Development Center, Indian Institute of Technology Kharagpur, Kharagpur, India-721302*

*[2]Department of Chemical Engineering, Indian Institute of Technology Kharagpur, Kharagpur, India-721302*

*[3]Department of Mechanical Engineering, Indian Institute of Technology Kharagpur, Kharagpur, India-721302*

The electroosmotic flow of non-Newtonian fluids in deformable microchannels is fundamentally important in the understanding of the hydrodynamics in physiological flows. The performance of these microchannels is governed by the load bearing capacity indicating the maximum amount of load that the device can withstand. While significant research efforts are aimed towards the coupling of electrokinetics with substrate deformability, the corresponding enhancement in the performances still remains elusive. Towards this, employing an intricate coupling between substrate compliance, hydrodynamic, and electrokinetic modulations, we have analyzed the possible sources of alterations in the flow physics in a deformable microchannel under the rheological premises of viscoelastic fluids which have a close resemblance with biological fluids typically used in several bio and micro-fluidic applications. The present study reveals that by operating under favorable regimes of parameters like the concentration and molecular weight of the polymer, the quality of the Newtonian solvent, and the concentration of electrolyte, one can achieve substantial augmentation in the load carrying capacity of a deformable microchannel for viscoelastic fluids as compared to its Newtonian counterpart. We believe that the present theoretical framework can be extremely important in the designing of electro-kinetically modulated bio-mimetic microfluidic devices.

* E-mail address for correspondence: suman@mech.iitkgp.ac.in

## 1. Introduction

Fluid-structure interaction (FSI) is one of the most interesting domains of modern-day microfluidic research which is originated from the deformability of a microfluidic channel [1–4]. In FSI problems, the deformation of the microchannel manifested in terms of a perturbed pressure distribution influences the hydrodynamics within the microchannel which in turn alters the resulting degree of deformation thereby establishing a fascinating two-way coupling under a microfluidic environment [5–8]. Accordingly, significant research contributions are noticeable in recent years demonstrating the interplay of electro-mechanics, structural mechanics, and fluid mechanics over small scales [9–12].

Most of the research works under the realm of electroosmotic flow in deformable microchannels are based on the assumption of uniform surface potential and slip. However, these microfluidic channels are often characterized by wettability and surface charge modulations [13–17], typically observed in narrow confinement flows interfacing with biological premises. Also, patterned variations of wettability or surface charge can be induced by the virtue of the engineered approach [18–22]. Since these variations of surface potential and slip are inevitable in practical scenarios, these modulations should be incorporated in the modeling of electroosmotic flow in deformable microchannels.

Another crucial aspect is the significance of the substrate deformability and its implications on the resulting flow dynamics [3,4,23,24]. While it may not be that significant in certain flow conditions, it becomes critically important in FSI problems. Interestingly, a major portion of the experimental microfluidic research is based on polymeric materials like Poly dimethyl siloxane (PDMS) and Polymethyl methacrylate (PMMA) (involved in the micro-fabrication process) which are inherently deformable [25–30]. In general, all these aspects are important in the modeling of electroosmotic flow in deformable microchannels. While some of these aspects are addressed earlier individually, these are interrelated and are needed to be brought into a common theoretical framework.

Fundamentally, the variations in surface potential and slip break the uni-directionality of a flow field accompanied by the generation of a secondary flow component and a perturbed pressure distribution. This subsequently tries to deform the microchannel which is resisted through the stress response stemming from the compliant nature of the microchannel. This stress response further influences the pressure distribution through the normal stress balance at the solid-fluid interface thus directly influencing the load bearing performance of the microchannel. In lubricated systems [3,4,31–37], load bearing capacity is the quantification of the enduring maximum amount of load [3,4,38–43]. While an effort to couple interfacial phenomenon like electrokinetics with surface compliance is noticeable in recent years [9,29,44–48], comprehensive theoretical modeling of the electroosmotic flow in deformable microchannels considering all aforesaid aspects is lacking, let alone the possible several alterations in the hydrodynamics under the rheological premises of a complex fluid. A propensity of deploying complex non-Newtonian fluids can be noticed in recent years in several micro and nanofluidic applications because of their close resemblances with biological fluids [49–53]. Subsequently, the rheological characteristics of these biofluids are modelled using a wide range of constitutive relationships, namely, power-law, [54] Casson, [55] Carreau models, [56,57] for inelastic fluids and Maxwell, [53,58] Oldroyd-B, [59] Phan-Thien Tanner models [60–65] for viscoelastic fluids.

Considering all these aforementioned aspects, we have theoretically studied the electroosmotic flow of a viscoelastic fluid through a parallel plate deformable microchannel subjected to the axial modulations of surface potential and slip. Appealing to the practical scenarios, the present analysis shows that by making a suitable combination of experimentally tunable parameters like polymer concentration, polymer molecular weight, regulating the quality of Newtonian solvent, and altering the electrolyte concentration, one can realize a significant augmentation in the load bearing capacity of a deformable microfluidic channel. We envisage that the present study holds practical relevance in constructing a new paradigm towards novel design and optimal performance of bio-mimetic lubricated devices where the fluidic, elastic, electrical, and rheological parameters can be coupled together over small scales.

## 2. Problem Formulation

The physical domain of the present analysis consists of a parallel plate deformable microchannel where Figs. 1(i) and (ii) depicts the un-deformed and deformed configurations respectively. A rectangular Cartesian co-ordinate system is chosen where $x$ and $y$ are the longitudinal and transverse co-ordinates with $y = 0$ being the origin of the physical domain. The channel ends are in isobaric and isothermal conditions. The length of the microchannel $(L)$ is much higher as compared to the half channel height, $(H)$ i.e., $L >> H$, so, $H/L << 1$, i.e., the lubrication approximation can be considered. Here, the flow is actuated using an axially applied electric potential where a binary, symmetric (1:1) electrolyte solution is employed. The channel walls are subjected to axial modulation in the hydrodynamic slip length which takes the form $l_s\left[1+\delta\cos\left(q\,x+\theta\right)\right]$ where $l_s$ is the slip length, $\delta$ amplitude of axial modulation, $q$ patterning frequency, and $\theta$ phase difference between the axially varying and invariant components. Interestingly, some research works highlighted the significance of hydrodynamic slip in altering the surface potential by following a functional dependence $\zeta_{eff}=\zeta\left(1+\kappa\, l_{s_{eff}}\right)$ [66–68] where $\zeta$ is the reference surface potential of hydrophilic surface and $\kappa$ is the inverse of electrical double layer (EDL) thickness. Since the slip length is axially modulated, surface potentials are also subjected to similar axial variation: $\zeta_{eff}=\zeta\left(1+\kappa\, l_{s_{eff}}\right)=\zeta\left[1+\kappa\, l_s\,\delta\cos\left(q\,x+\theta\right)\right]$.

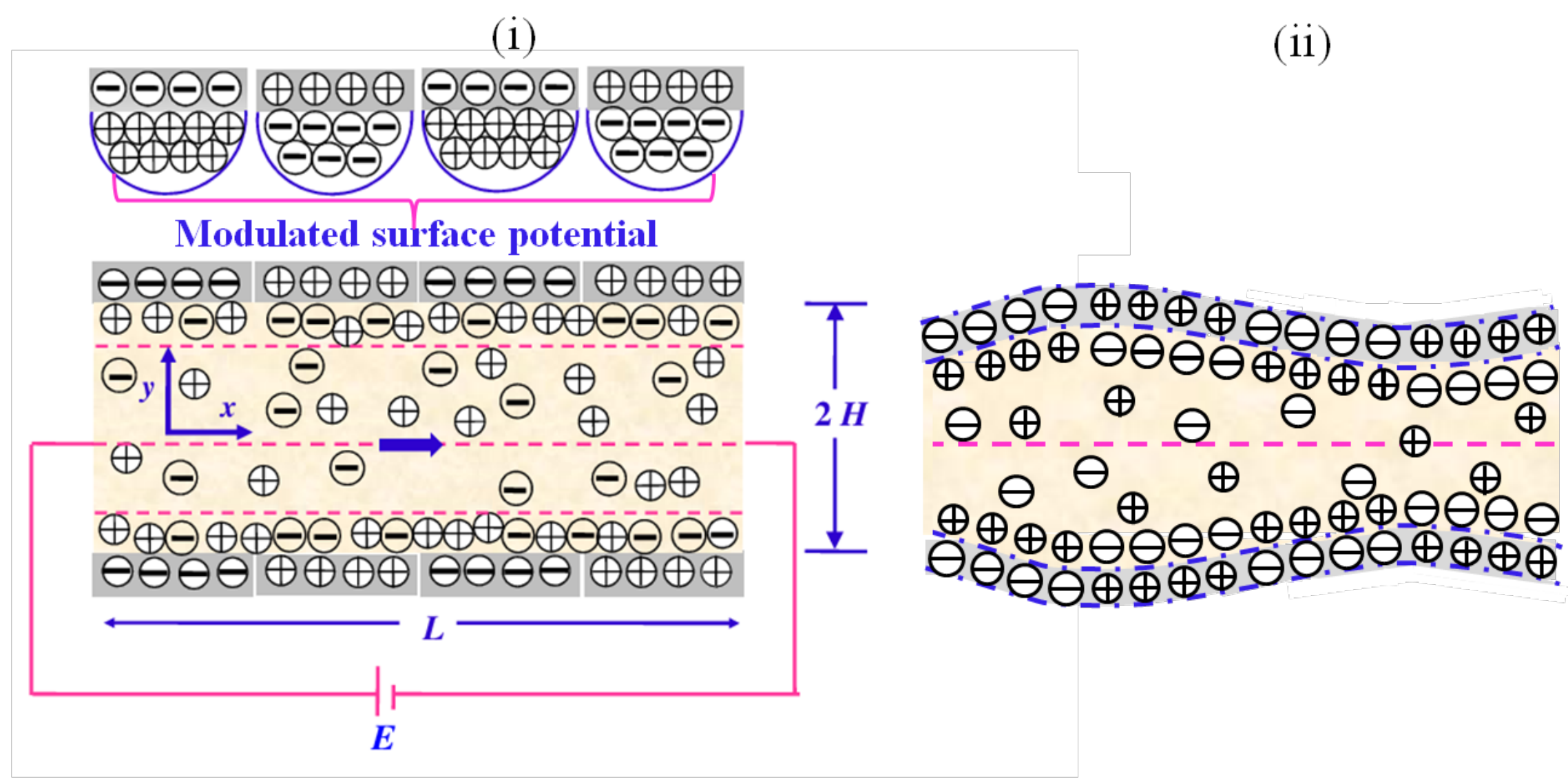

**Fig. 1.** Schematic of electroosmotic flow through a parallel plate microchannel (i) undeformed configuration, (ii) deformed configuration.

In absence of hydrodynamic and electrokinetic modulations, it is purely electroosmotic flow with a uniform velocity profile and therefore, the channel deformability does not affect the flow field. Any perturbation comes from the modulations of slip and surface potential results in an imbalance in the pressure distribution. Now the channel walls, because of compliance, tend to maintain the original configuration. This in turn perturbs the pressure field and the microchannel undergoes finite deformation. The degree of deformation is obtained using the equation of equilibrium for the stress field in the deformable wall [4,69,70]

$$\nabla \cdot \sigma = 0 \tag{1}$$

where $\sigma = \vartheta\left(\nabla\vec{h} + \nabla\vec{h}^T\right) + \Lambda\nabla\cdot\vec{h}[I]$ is the Cauchy-Green stress tensor. For an isotropic, linearly elastic solid substrate, $\vartheta$ and $\Lambda$ are the Lamé constants while $\vec{H} = [H_x, H_y]^T$ is the displacement field vector. With shallow elastic layer consideration, Eq. (1) is simplified to the following form

$$\frac{\partial^2 H_y}{\partial y^2} = 0 \tag{2}$$

To solve this equation, one needs to incorporate the no-displacement condition at the complaint layer-solid interface (i.e., at $y = H$, $h_y = 0$) while normal stress balance prevails at the complaint layer-fluid interface (i.e., at $y = H - H_c$, $\sigma\cdot\hat{n} = -p\,\hat{n}$). The resulting simplified form of $H_y$ is

$$H_y = -\left(\frac{H_c}{2\vartheta + \Lambda}\right)p \tag{3}$$

In Eq. (3), $p$ is the hydrodynamic pressure that relates the elastic domain to the fluidic domain. The compliant layer thickness is taken as $H_c$ which correlates the stiffness factor of the channel as $\left(\frac{H_c}{2\vartheta + \Lambda}\right) = S^{-1}$ where $S$ is the stiffness of the channel. Depending on $S$, one can experience a transition from the compliant regime to the stiff regime.

For the transport phenomenon, the flow is assumed to be steady, laminar, incompressible, and in the creeping flow regime. The governing continuity and momentum equations are

$$
\begin{array}{ll}
\text{Continuity Equation:} & \nabla \cdot \boldsymbol{v} = 0 \\
\text{Momentum Equations:} & \begin{cases} x\text{-component}: & 0 = -\dfrac{\partial p}{\partial x} + \dfrac{\partial \tau_{xx}}{\partial x} + \dfrac{\partial \tau_{yx}}{\partial y} + \rho_e E \\ y\text{-component}: & 0 = -\dfrac{\partial p}{\partial y} + \dfrac{\partial \tau_{xy}}{\partial x} + \dfrac{\partial \tau_{yy}}{\partial y} \end{cases}
\end{array}
\tag{4}
$$

where $\boldsymbol{\tau}$ is the stress tensor, $E = -\nabla\phi$ induced electric field due to the application of the electric potential $(\phi)$, $\rho_e$ excess change density. Now, the excess charge density is obtained from the Poisson equation $\nabla^2\phi = -\rho_e/\varepsilon$ with $\varepsilon$ being the electrical permittivity of the fluid. Considering weak field approximation, $\phi$ is expressed as $\phi(x,y) = \phi_{ext}(x) + \psi(x,y)$ with $\phi_{ext}(x)$ and $\psi(x,y)$ being the externally imposed and induced (within EDL) electric potentials respectively. The Boltzmann distribution assumption for the ionic number density along with the lubrication approximation results $\nabla^2\phi \approx d^2\psi/dy^2 = \kappa^2 \sinh(\psi)$. Invoking Debye-Hückel linearization approximation [14,71,72], this is further simplified as $d^2\psi/dy^2 = \kappa^2\psi$. Using $\overline{x} = x/L$, $\overline{y} = y/H$, $\overline{p} = (p - p_{atm})H^2/\mu u_c L$, $\overline{\tau}_{ii} = \tau_{ii} H^2/\mu u_c L$, $\overline{\kappa} = \kappa H$, $\overline{l_s} = l_s/H$, $\overline{\psi} = \psi/\zeta$ and $\overline{u} = u/u_c = u/u_{hs} = u/(-\varepsilon E\zeta/\mu)$ ($u_c = u_{hs} = -\varepsilon E\zeta/\mu$ is the characteristic velocity scale, commonly known as Helmholtz-Smoluchowski velocity scale), the dimensionless form of the momentum equation reads as

$$
0 = -\frac{d\overline{p}}{d\overline{x}} + \frac{\partial \overline{\tau}_{xy}}{\partial \overline{y}} + \overline{\kappa}^2\,\overline{\psi} \quad (x - \text{component}) \tag{5}
$$

where $\partial\overline{p}/\partial\overline{x}$ is replaced by $d\overline{p}/d\overline{x}$ since $\partial\overline{p}/\partial\overline{y} = 0$. Additionally, the axial electric field no longer remains uniform in a deformable microchannel which becomes axially variant to maintain the current conservation [73]. With the electrical conductivity ($\sigma$) of fluid remaining spatially invariant, the simplified form of the current conservation can be written as $\dfrac{\partial}{\partial\overline{x}}\left\{\overline{E}(\overline{x})\,\overline{H}(\overline{x})\right\} = 0$.

To describe the rheological behavior of a viscoelastic fluid, we have chosen the constitutive equation of the simplified Phan-Thien-Tanner model (sPTT) [60–63,74–76]. For this model, the stress components take the following form [64,77,78]

$$F\tau_{xx}+\lambda\left(u\frac{\partial\tau_{xx}}{\partial x}+v\frac{\partial\tau_{xx}}{\partial y}-2\frac{\partial u}{\partial x}\tau_{xx}-2\frac{\partial u}{\partial y}\tau_{yx}\right)=2\mu\frac{\partial u}{\partial x}$$
$$F\tau_{yy}+\lambda\left(u\frac{\partial\tau_{yy}}{\partial x}+v\frac{\partial\tau_{yy}}{\partial y}-2\frac{\partial v}{\partial x}\tau_{xy}-2\frac{\partial v}{\partial y}\tau_{yy}\right)=2\mu\frac{\partial v}{\partial y} \quad (6)$$
$$F\tau_{xy}+\lambda\left(u\frac{\partial\tau_{xy}}{\partial x}+v\frac{\partial\tau_{xy}}{\partial y}-\frac{\partial u}{\partial y}\tau_{yy}-\frac{\partial v}{\partial x}\tau_{xx}\right)=\mu\left(\frac{\partial u}{\partial y}+\frac{\partial v}{\partial x}\right)$$

where $\mu=\mu_P+\mu_S$ is the fluid viscosity consisting of both polymeric component $(\mu_P)$ and its Newtonian counterpart $(\mu_S)$ (i.e., solvent); $\lambda$ is the fluid relaxation time. $F$ is the stress coefficient $F=\left[1+s\lambda\, tr(\boldsymbol{\tau})/\mu\right]$ with *s* being the fluid extensibility which in the limit of $s\to 0$ represents a simpler upper-convected Maxwell (UCM) model. Using the above-mentioned scales for the stresses along with lubrication approximation, the relationship between the velocity gradient and the stress tensor is simplified which is given below

$$\frac{\partial\overline{u}}{\partial\overline{y}}=\overline{\tau}_{xy}+2\in\mathrm{De}^2\overline{\tau}_{xy}{}^3 \quad (7)$$

where $De=\lambda\kappa u_{hs}$ is Deborah number representing the relative strength of elastic and viscous forces. The governing equations are subjected to modulated slip and surface potential boundary conditions as stated earlier. The solution procedure is discussed in the following section.

## 3. Semi-analytical solution procedure

The governing equations are analytically intractable and hence, no closed-form solution is possible. Therefore, an asymptotic approach is followed utilizing regular perturbation technique where any variable $\chi$ is expanded as

$$\chi=\chi_0+\delta\,\chi_1+\delta^2\chi_2+.... \quad (8)$$

where $\delta$ is the perturbation parameter with '0', '1', '2' superscripts representing leading-order, first-order, and higher-orders respectively. Now, the governing equations for O(1) flow field are

$$\left.\begin{aligned}&\frac{\partial\overline{\tau}_{xy,0}}{\partial y}=\overline{\kappa}^2\overline{\psi}_0\\&\frac{\partial\overline{u}_0}{\partial\overline{y}}=\overline{\tau}_{xy,0}+2\in\mathrm{De}^2\overline{\tau}_{xy,0}{}^3\end{aligned}\right\} \quad (9)$$

where $\overline{\psi}_0 = \cosh\left(\overline{\kappa}\,\overline{y}\right)\big/\cosh\left(\overline{\kappa}\overline{H}\right)$. Eq. (9) is subjected to constant slip length and surface potential boundary conditions. Since, the $\mathrm{O}\left(1\right)$ solution refers to perturbation free flow field, the effect of surface deformability is absent. The corresponding velocity profile is as follows

$$\overline{u}_0 = \left[\left\{1-\frac{\cosh\left(\overline{\kappa}\,\overline{y}\right)}{\cosh\left(\overline{\kappa}\overline{H}\right)}\right\}+\overline{l}_s\,\overline{\kappa}\tanh\left(\overline{\kappa}\overline{H}\right)\right]\left(1+\overline{\kappa}\,\overline{l}_s\right)+\frac{s\,De^2\left(1+\overline{\kappa}\,\overline{l}_s\right)^3}{2\cosh^3\left(\overline{\kappa}\overline{H}\right)}\begin{bmatrix}\dfrac{1}{3}\left\{\cosh\left(3\overline{\kappa}\overline{H}\right)-\cosh\left(3\overline{\kappa}\,\overline{y}\right)\right\}\\ -3\left\{\cosh\left(\overline{\kappa}\overline{H}\right)-\cosh\left(\overline{\kappa}\,\overline{y}\right)\right\}\\ +4\,\overline{\kappa}\,\overline{l}_s\sinh^3\left(\overline{\kappa}\overline{H}\right)\end{bmatrix} \tag{10}$$

Eq. (10) represents the electroosmotic flow of viscoelastic fluids in a slit microchannel with constant interfacial slip length. In the limit of $De \to 0$ and $\overline{l}_s \to 0$, this simplifies to the purely electroosmotic flow of a Newtonian fluid $\overline{u}_0 = \left[1-\left\{\cosh\left(\overline{\kappa}\,\overline{y}\right)\big/\cosh\left(\overline{\kappa}\overline{H}\right)\right\}\right]$). Similarly proceeding, the $\mathrm{O}\left(\delta\right)$ velocity profile is given by

$$\begin{aligned}\overline{u}_1 = {} & \frac{1}{2}\frac{d\overline{p}_1}{d\overline{x}}\left(\overline{y}^2-\overline{H}^2-2\,\overline{l}_s\,\overline{H}-12\,a_{64}\,\overline{l}_s\,\overline{H}\,{a_{42}}^2\right)+\frac{3}{2}\frac{a_{63}\left(\dfrac{d\overline{p}_1}{d\overline{x}}\right)}{\overline{\kappa}^2\cosh^2\left(\overline{\kappa}\,\overline{H}\right)}\left\{\begin{matrix}\overline{\kappa}\,\overline{y}\sinh\left(2\overline{\kappa}\,\overline{y}\right)-\overline{\kappa}\overline{H}\sinh\left(2\overline{\kappa}\,\overline{H}\right)\\ -\overline{\kappa}^2\left(\overline{y}^2-\overline{H}^2\right)-\cosh^2\left(\overline{\kappa}\,\overline{y}\right)+\cosh^2\left(\overline{\kappa}\,\overline{H}\right)\end{matrix}\right\}\\ & +\left[\begin{matrix}\dfrac{\overline{l}_s\,a_{44}\,\overline{\kappa}a_{42}}{\overline{H}}\left\{1+2\,a_{64}\,\overline{\kappa}\,{a_{42}}^2\right\}+\dfrac{a_{52}\,{a_{45}}^2}{\overline{H}}\left\{1-\dfrac{\cosh\left(\overline{Q}\,\overline{y}\right)}{\cosh\left(\overline{Q}\,\overline{H}\right)}\right\}+\dfrac{a_{45}\,{a_{52}}^2a_{41}}{\overline{H}}\left(1+6\,a_{64}\,{a_{42}}^2\right)\\ -\dfrac{3\,a_{64}\,a_{46}\,a_{52}}{2\cosh^2\left(\overline{\kappa}\,\overline{H}\right)\cosh\left(\overline{Q}\,\overline{H}\right)\overline{H}}\begin{bmatrix}\left(1/a_{55}\right)\left\{\cosh\left(a_{55}\,\overline{y}\right)-\cosh\left(a_{55}\,\overline{H}\right)\right\}\\ +\left(4/a_{56}\right)\left\{\cosh\left(a_{56}\,\overline{y}\right)-\cosh\left(a_{56}\overline{H}\right)\right\}\\ -\left(2/\overline{Q}\right)\left\{\cosh\left(\overline{Q}\,\overline{y}\right)-\cosh\left(\overline{Q}\,\overline{H}\right)\right\}\end{bmatrix}\end{matrix}\right]\cos\left(\overline{q}\,\overline{x}+\overline{\theta}\right)\end{aligned} \tag{11}$$

The coefficients of Eq. (11) are reported in **Section A** of the **Appendix**. Unlike $\mathrm{O}\left(1\right)$ solution, here, the channel height is axially variant $\overline{H}\left(\overline{x}\right)$ which incorporates the effect of surface compliance as $\overline{H}\left(\overline{x}\right)=\overline{h}+\overline{d}\left(\overline{x}\right)$ where $\overline{h}$ is the axially invariant part and $\overline{d}\left(\overline{x}\right)=\beta'\,\overline{p}_1\left(\overline{x}\right)$ is the deformation with $\beta' = \mu\,u_c\,L\big/\left(S\,H^3\right)$. Now, the two-dimensional flow is represented in terms of classical Reynolds equation as [4,70]

$$\int_{-\overline{H}}^{\overline{H}}\frac{\partial\overline{u}_1}{\partial\overline{x}}\,d\overline{y}+\int_{-\overline{H}}^{\overline{H}}\frac{\partial\overline{v}_1}{\partial\overline{y}}\,d\overline{y}=0 \tag{12}$$

where the $v$-component can be obtained by using the no-penetration condition at the surface (i.e., $\bar{v}\big|_{\bar{y}=\pm\bar{H}}=0$). Further simplification of Eq. (12) is done by applying Leibniz's rule of differentiation under the integral which yields

$$\int_{-\bar{H}}^{\bar{H}}\frac{\partial\bar{u}_1}{\partial\bar{x}}d\bar{y}=\frac{\partial}{\partial\bar{x}}\int_{-\bar{H}}^{\bar{H}}\bar{u}_1 d\bar{y}-\bar{u}_1\left(\bar{y}=\bar{H}\right)\frac{\partial\bar{H}}{\partial\bar{x}}+\bar{u}_1\left(\bar{y}=-\bar{H}\right)\frac{\partial\left(-\bar{H}\right)}{\partial\bar{x}}=0 \tag{13}$$

Eq. (13) results in a differential equation describing the following pressure distribution in the axial direction where some mathematical simplifications are chosen like $\left\{\bar{H}\left(\bar{x}\right)\right\}^3\approx\bar{h}^3$, $\left\{\bar{H}\left(\bar{x}\right)\right\}^2\approx\bar{h}^2$ and $\tanh\left\{\bar{\kappa}\,\bar{H}\left(\bar{x}\right)\right\}\approx\tanh\left(\bar{\kappa}\,\bar{h}\right)$. After some rescaling $\tilde{x}=x/L,\ \tilde{y}=\tilde{H}=H/L, \tilde{\kappa}=\kappa L,$ $\tilde{p}=\left(p-p_{atm}\right)H/\mu\, u_c$, $\tilde{q}=q\,L,\ \tilde{l}_s=l_s/L,\ \tilde{\psi}=\psi/\zeta,\ \beta=\mu\, u_c/S\,L^2,$ the final form of the governing equation for $\mathrm{O}\left(\delta\right)$ pressure distribution is given by

$$\chi_1\frac{d^2\tilde{p}_1}{d\tilde{x}^2}+\chi_2\left(\frac{d\tilde{p}_1}{d\tilde{x}}\right)^2+\chi_3\frac{d\tilde{p}_1}{d\tilde{x}}+\chi_4=0 \tag{14}$$

Eq.(14) involves a non-dimensional parameter $\beta=\mu\, u_c/S\,L^2$ which signifies the relative contribution of the viscous effect to the elastic effect with parameter $S$ indicating the effective stiffness per unit area. This parameter arises from the coupling of fluid and solid domains. Using typical values of the pertinent variables (these are reported in **Table 1**) one can realize that $\beta$ is a small quantity, so, $\beta^2 << 1$. Accordingly, any term involving $\beta^2$ is neglected; and O($\beta^0$) and O($\beta$) terms are retained before arriving on Eq. (14). The coefficients involved in Eq.(14) can be found in **Section A** of the **Appendix**. Since Eq. (14) is non-linear, it is not analytically tractable. Hence, a numerical technique like the shooting method is employed to obtain the pressure distribution subjected to the isobaric condition at the two exits (i.e., at $x=0,\ l;\ p=p_{atm}$). Then, the pressure distribution can be used to evaluate the load bearing capacity of the channel $\left(\tilde{W}\right)$ as

$$\tilde{W}_1=\frac{W_1}{\mu\, u_c}=\delta\int_0^1\bar{p}_1\,d\bar{x} \tag{15}$$

For sake of conciseness, the governing equations for higher order are not included.

## 4. Results and Discussions

Since a large number of variables are involved in this analysis, one can obtain numerous results depending on their parametric variations. However, we here report only some specific case studies consistent with physically realizable scenarios such that one can draw a perspective from an experimental point of view. Towards this, some typical values of the parameters are chosen which are listed below in **Table 1**:

| **Parameters** | **Physical significance** | **Typical values** |
|---|---|---|
| $L$ | longitudinal length scale | 1 mm |
| $H$ | transverse length scale | 10 $\mu$m |
| $E_{\text{ref}}$ | reference electric field strength | $10^4$ V/m |
| $l_s$ | slip length | 1-10 nm |
| $\mu_{\text{ref}}$ | reference fluid viscosity | $10^{-3}$ Pa.s. |
| $\lambda_D$ | EDL thickness | 10 nm |
| $\theta$ | phase-shift angle | $0^{\circ}$ (i.e., no phase-shift angle) |
| $q$ | patterning frequency | $2\pi/L$ $\text{m}^{-1}$ |
| $S$ | stiffness per unit area | $10^6$ N/m$^3$ |

**Table 1**. List of parameters involved along with their physical significances and typical values.

In this context, it is worthwhile to mention that Deborah number (De) is a very important characteristic dimensionless number associated with the flow of viscoelastic fluids representing the relative strength between elastic and viscous forces within the flow domain. Intuitively, one can be keen to see its effect on the flow field and the associated deformation characteristics. Accordingly, the augmentation of the load bearing capacity in viscoelastic fluids $\left(\tilde{W}\right)$ is plotted as a function of Deborah number (De) in Fig. 2(a) for two different slip lengths 1 nm (i.e., $\tilde{l}_s$ = 1) and 10 nm (i.e., $\tilde{l}_s$ = 10) respectively. Here, De = 0 refers to the Newtonian fluid and $\tilde{W}_{\text{ref}}$ is the reference load bearing capacity corresponding to the Newtonian fluid. Irrespective of the slip length, the variation of $\tilde{W}/\tilde{W}_{\text{ref}}$ with Deborah number follows a power-law type dependence (i.e., $\tilde{W}/\tilde{W}_{\text{ref}}$ ~ $\text{De}^{1.85}$ and $\text{De}^{3.74}$). For both slip lengths, at lower De, similar dependence of load bearing capacity on De is observed where the plots are identical up to De = 0.3 beyond which the degree of non-linearity in the variation amplifies significantly at higher slip length (i.e., $l_s$ = 10 nm). Increasing De from 0 to 1 result in an augmentation of ~ 2.65 times in the load bearing capacity for lower slip length (i.e., $l_s$ = 1 nm) while the corresponding enhancement at higher slip length (i.e., $l_s$ = 10 nm) is ~ 5.25 times as can be seen from Fig. 2(a).

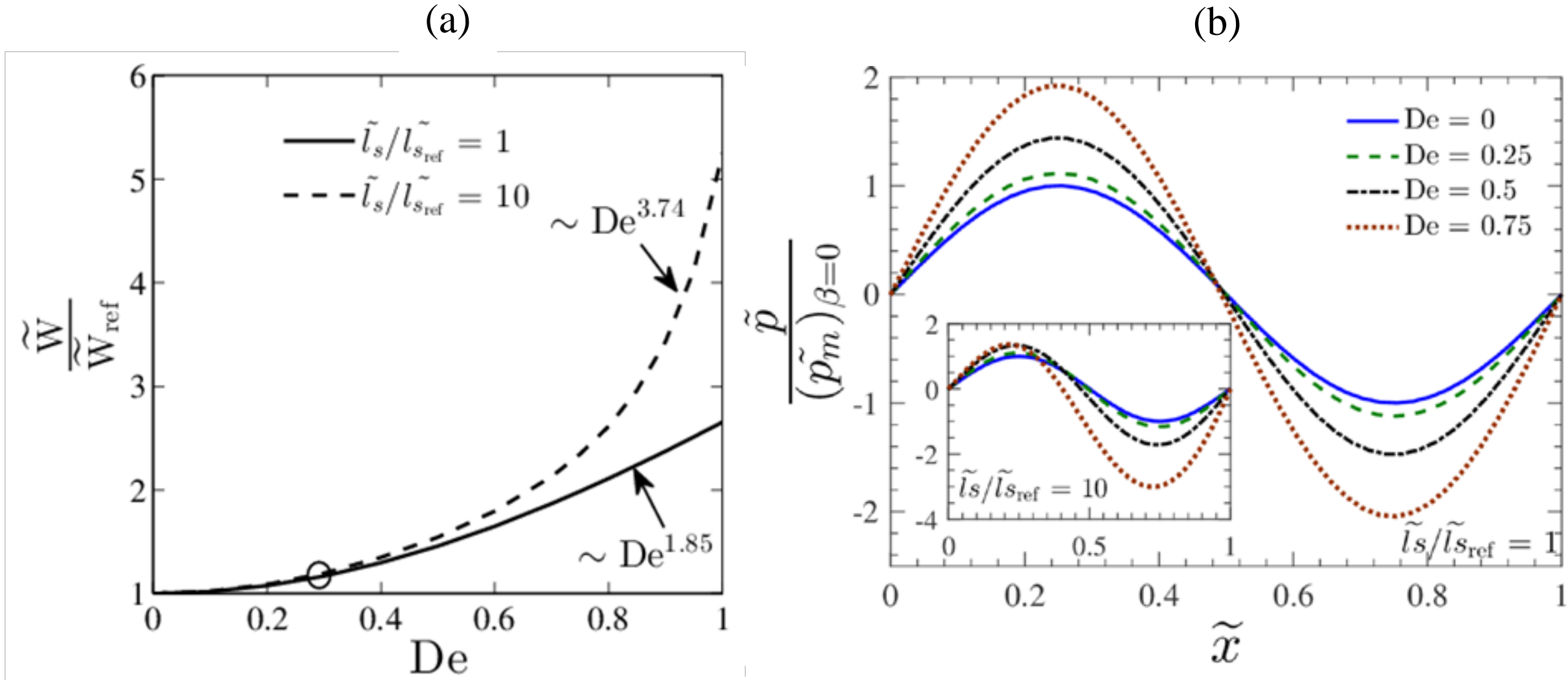

**Fig. 2.** (a) Load bearing capacity $\left(\tilde{W}/\tilde{W}_{\mathrm{ref}}\right)$ variation with Deborah number (De) for two different slip lengths ($l_s$) 1 nm and 10 nm respectively where $l_{s_{ref}}$ is 1 nm. (b) The pressure distribution in the *x*-direction for different values of De evaluated at $l_s$ = 1 nm. Inset shows the corresponding results for $l_s$ = 10 nm.

Since the genesis of the load bearing capacity is the perturbed pressure profiles, the corresponding pressure distributions in the axial direction are depicted in Fig. 2(ii). Here, the pressure profiles are normalized using a variable $\tilde{p}_m$ where $\tilde{p}_m$ indicates the maximum value of the magnitude of pressure distribution in the case of a Newtonian fluid. Physically, the flow is significantly modulated already under axial variations of slip and surface potential and the compliance nature of surface establishes solid-fluid interaction. Now, for a viscoelastic fluid, an additional source of alteration comes into play, i.e., the implications of elastic stresses. Interestingly, similar to the effects of slip and surface potential modulations, the elastic stresses are also localized in the wall-adjacent region. Therefore, the additional source of disturbance due to viscoelasticity-induced perturbation is manifested only in the near-wall region. This effect in association with the slip and surface potential modulations results in a strong perturbation to the flow field responded by the channel flexibility. This in turn generates a stronger perturbed pressure field as reflected in the pressure profiles in Fig. 2(b) with different Deborah numbers. Now if one compares Fig. 2(b) with its inset, the effect of slip length can be realized. At lower slip length (i.e., $l_s$ = 1 nm), pressure profiles are similar (of sinusoidal nature) with maxima occurring at the downstream and minima at the upstream. At higher slip length (inset of Fig. 2(b)), however, more is the alteration in the pressure distributions with increasing Deborah number indicating a stronger influence of viscoelasticity in governing the flow dynamics and accordingly, the same is reflected in the variations of load bearing capacities in Fig. 2(a).

After realizing the possible outcome of the parametric variation of Deborah number (De), we look into the definition of the same, i.e., $\mathrm{De} = \lambda \kappa u_c$. It clearly shows the dependence on two physical properties, i.e., fluid viscosity $(\mu)$ and fluid relaxation time $(\lambda)$. Interestingly, both of these two properties can easily be modulated employing the polymer concentration $(c)$. Therefore, it would be more realistic to explore the possible sources of alteration in the load bearing capacity using polymer concentration so that it can be executed in actual experimental scenarios. As reported earlier, depending on the polymer concentration $(c)$, a solution can experience a transition from one regime to another (regarding this a discussion is included in **Section B** of the **Appendix**) [79–84]. These regimes are divided into three parts, namely, dilute, semi-dilute unentangled, and semi-dilute entangled. Here, two representative examples of the aqueous solution of Polyethylene Oxide (PEO) and Polyacrylamide (PAM) [79] are chosen where different regimes are demarcated by I, II, and III in Figs. 3(a)-(b) with $c^*$ and $c_e$ being the overlap and entanglement concentrations respectively. In the dilute regime, as $c$ is increased gradually, as reflected in the scaling relationship discussed in **Section B** of the **Appendix**, viscosity $(\mu)$ follows a linear relationship with $c$ while relaxation time $(\lambda)$ remains constant for a dilute solution (for a dilute solution, Zimm's relaxation time $(\lambda_z)$ is typically employed which is independent of $c$, i.e. $\lambda \propto c^0$). At a very low polymer concentration, i.e., at $c$ = 0.06, the viscosity is relatively lower which also possesses a finite degree of elasticity. As far as the enhancement of the load bearing capacity using viscoelastic fluid is concerned, the elasticity-mediated disturbance is advantageous creating a deformation of the polymer chain away from its equilibrium position whereas the viscous effect is detrimental where the viscous resistance suppresses the hydrodynamic or rheological perturbations. Therefore, at lower polymer concentration, the elastic effect dominates over viscous effect resulting in an increment of ~ 1.43 times in the load capacity ratio at $c$ = 0.06. With increasing polymer concentration ($c$), higher is the significance of the viscous resistance-induced flow suppression resulting in a decay in the load capacity ratio. As shown in insets (i), (iii) of Fig. 3, a rise of ~ 1.4 times in viscosity is observed. Accordingly, the load capacity ratio $\left(\tilde{W}/\tilde{W}_{\mathrm{ref}}\right)$ undergoes a reduction from ~ 1.43 to ~ 1.23 in PEO (or ~ 1.21 in PAM solution) solution as the polymer concentration varies from $c$ = 0.06 to $c$* (i.e., overlap concentration). This reduction follows a power-law decay type of relationship: $\tilde{W}/\tilde{W}_{\mathrm{ref}} \sim c^{-0.24}$ (for both PEO and PAM solutions).

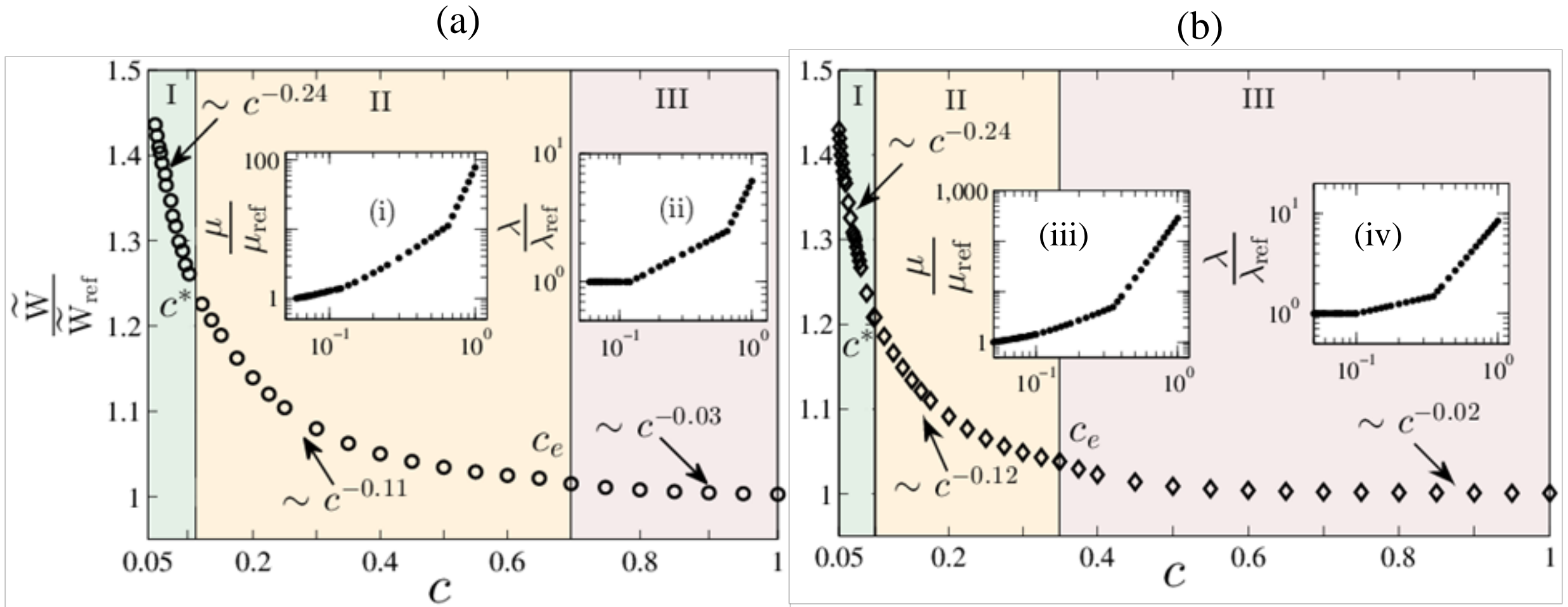

**Fig. 3.** Load bearing capacity $\left(\tilde{W}/\tilde{W}_{\mathrm{ref}}\right)$ variation with increasing concentration of the aqueous polymer solution $(c)$. (a) refers to aqueous Polyethylene Oxide (PEO) solution while (b) is for aqueous Polyacrylamide (PAM) solution. Also, insets (i), (iii) are the results for viscosity variation while inset (ii), (iv) corresponds to relaxation time $(\lambda)$ variation with $c$. $c^*$ and $c_e$ are the overlap and entanglement concentrations respectively (for both cases). $\tilde{W}_{\mathrm{ref}}$ is the load bearing capacity for Newtonian fluids.

When one moves from dilute to semi-dilute unentangled regime, with increasing $c$, both fluid relaxation time ($\lambda$) and viscosity ($\mu$) increases. As shown in insets of Fig. 3(b), changing $c$ from $c^*$ (i.e., overlap concentration) to $c_e$ (i.e., entanglement concentration) results in an increase of ~ 1.5 times for $\lambda$ and ~ 3.5 times for $\mu$ in the semi-dilute regime. Physically, higher $\lambda$ enhances elastic effect creating more departure of the polymer chain from its equilibrium configuration while higher $\mu$ resists the same. Since the rate of amplification in the viscous resistance is higher compared to the elastic effect, it overshadows the latter thereby lessening the load capacity ratio significantly. Here, changing the polymer concentration from $c^*$ to $c_e$ lowers the load capacity ratio from ~ 1.23 to ~ 1.02 in PEO solution (or ~ 1.04 in PAM solution).

In the semi-dilute entangled regime, the dependence of $\lambda$ and $\mu$ on polymer concentration becomes stronger. For example, changing the polymer concentration ($c$) beyond $c_e$ in PEO solution results in an augmentation of $\mu$ and $\lambda$ up to ~ 77 times and ~ 6 times respectively. So, the rate of increment in $\mu$ becomes at least one order of magnitude higher as compared to that of $\lambda$. Therefore, despite the strengthened elasticity-induced disturbance, the strongly pronounced viscous effect dictates the hydrodynamics thus leading to a reduction in load capacity ratio ($\tilde{W}/\tilde{W}_{\text{ref}}$) beyond $c_e$. Increasing $c$ beyond $c_e$ creates a scenario when the ratio $\tilde{W}/\tilde{W}_{\text{ref}}$ approaches unity which implies that the load bearing capacity in viscoelastic fluid (for both PEO and PAM solutions) coincides with that of a Newtonian fluid. Hence, from an experimental point of view, if one wants to enhance the load bearing capacity of lubricated systems using viscoelastic fluids, solutions in the dilute regime should only be preferred over other regimes which also naturally involves only a small addition of polymer.

After establishing the dilute polymeric solution advantageous, we now look into the possible sources of rheological alteration of polymer solutions in the dilute regime. As per definition, a solution can be assumed as dilute when there is no interaction between the polymer chains [80,82,85–87]. Just a small addition of polymer initiates several thermodynamic interactions within a Newtonian solvent out of which hydrodynamic interaction is a significant one in which disturbance in the flow field is caused by the polymer chain at one part with the other part producing a drag force on the Newtonian solvent. This effect comes into prominence with the increasing molecular weight of the polymer $(M_w)$. Here, intrinsic viscosity $[\mu]$ is a parameter associated with a dilute solution which is utilized to determine the cross-over (or overlap) concentration ($c^*$) (a discussion about intrinsic viscosity is appended in **Section C** of the **Appendix**). According to Flory's theory, $c^*$ is the reciprocal of intrinsic viscosity $[\mu]$. Now, $[\mu]$ depends on $M_w$ as $[\mu] = a M_w^{\ b}$ (also known as Mark-Houwink-Sakurada (MHS) equation) [80] with *a* and *b* being experimentally obtained empirical constants. Physically, the higher the molecular weight ($M_w$), the lesser is the concentration required to reach the cross-over concentration. So, for higher $M_w$, very small concentration is strong enough to initiate interaction between the polymer chains. Additionally, the definition of Zimm's relaxation time $(\lambda_z)$ tells us that $\lambda_z$ is directly proportional to $M_w$. This aforesaid variations of $\lambda_z$ and $[\mu]$ are depicted in insets (i) and (ii) of Fig. 4 with dilute aqueous solution of PEO being chosen as an example. As clear from these figures, the influence of $M_w$ (in g/mol) on $\lambda_z$ is much stronger as compared to that on $[\mu]$ which enhances the elastic stresses significantly which in turn creates an amplified imbalance in pressure distribution. In the range $10^5 \le M_w \le 2\times10^6$ ($M_w$ in g/mol), the effect of molecular weight on the load capacity ratio ($\tilde{W}/\tilde{W}_{\text{ref}}$) is indistinguishable while beyond $M_w = 3\times10^6$ g/mol, it increases sharply up to ~ 5.8 times by following the relationship: $\tilde{W}/\tilde{W}_{\text{ref}} \propto M_w^{\ 3.68}$.

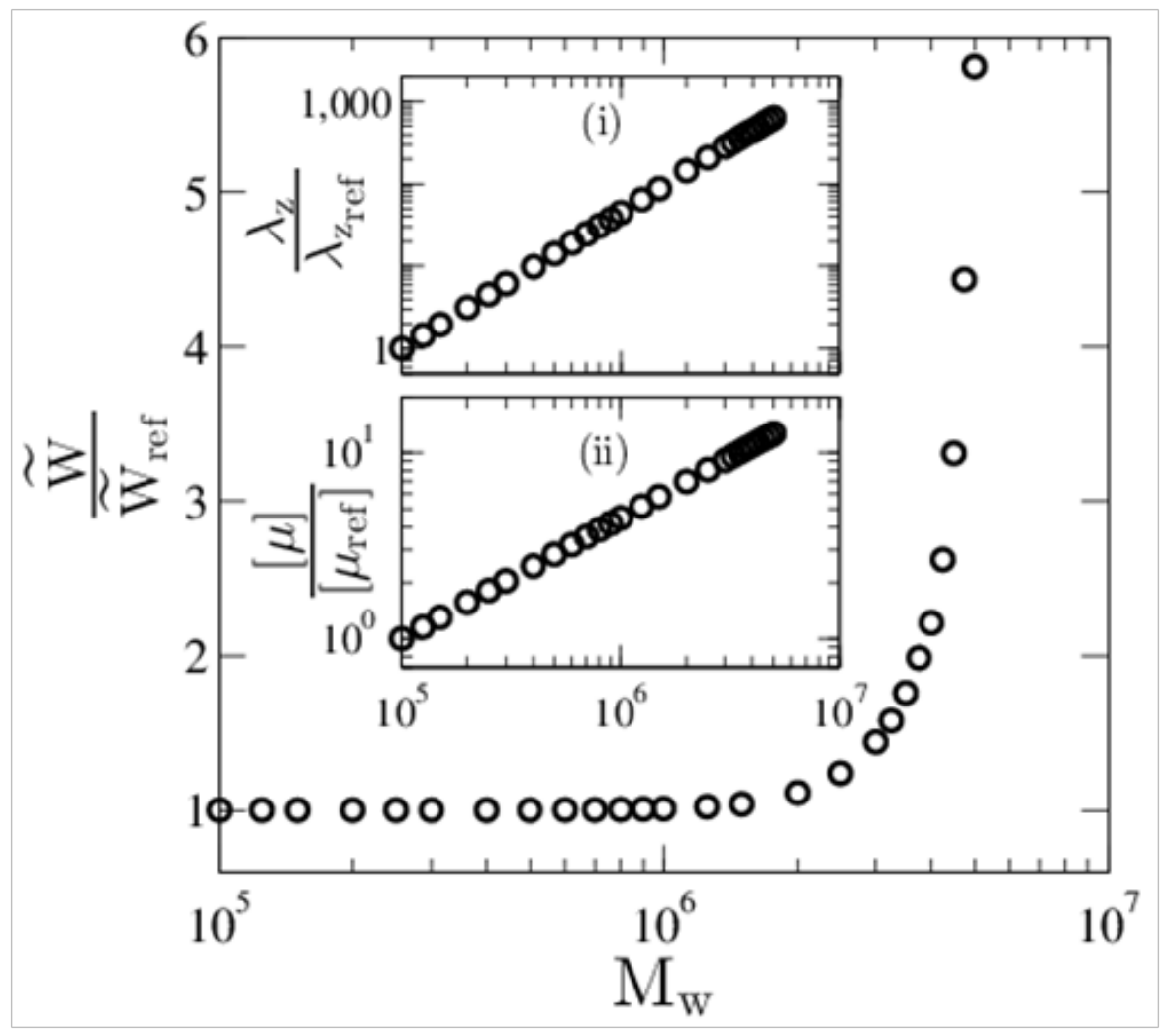

**Fig. 4.** The dependence of $\tilde{W}/\tilde{W}_{\text{ref}}$ as a function of polymer molecular weight $(M_w)$ for aqueous solution of polyethylene oxide (PEO). Inset (i) and (ii) are the variations of Zimm's relaxation time $(\lambda_z)$ and intrinsic viscosity of polymer $[\mu]$ with $M_w$.

Another feature is the effect of the excluded volume which becomes of critical importance when dealing with dilute polymer solution where the conformation of the polymer chain depends strongly on the quality of the Newtonian solvent. [79,87,88] The degree of expansion of the chain is dictated by the intermolecular interaction between the polymer molecules and the solvent molecules. This is associated with an energy of interaction that plays a pivotal role in determining the expansion or contraction of the chain. If the quality of the solvent is poor enough to nullify any effect of excluded volume, then it fulfills the $\theta$-condition thereby contracts the polymer. [89,90] On the contrary, in a good solvent, the repulsion between the chains is strong enough to expand the conformation. In Fig. 5(a), the effect of solvent quality on the load capacity ratio $\left(\tilde{W}/\tilde{W}_{\text{ref}}\right)$ is demonstrated as a function of polymer concentration in the dilute regime (i.e., $c$ is well below $c^*$). To illustrate this, two types of solvents are chosen as examples, Dioctyl Phthalate (DOP) and Tricresyl Phosphate (TCP), which are considered as near $\theta$-solvent (at $\theta$-temperature 22°C) and good solvents respectively. [88] The microfluidic approach of relaxation time determination in recent years have shown that by monitoring the solvent quality one can get significantly different values of relaxation time ($\lambda$) as compared to Zimm's relaxation time ($\lambda_z$). (regarding this, a brief discussion is added in **Section D** of the **Appendix**) Mathematically, for a fixed value of concentration ($c$) and molecular weight ($M_w$), intrinsic viscosity ($[\mu]$) is related as $[\mu] \propto M_w^{3\nu-1}$ which simplifies to $M_w^{0.5}$ for $\theta$-solvent and $M_w^{0.8}$ for good solvent respectively. Here $\nu$ signifies the solvent quality ranging from 0.5 ($\theta$-solvent) to 0.6 (good solvent). [79] Since the alteration in $[\mu]$ for $\theta$-solvent is less significant as compared to good solvent, the criterion for $c^*$ is achieved at higher $c$ thus resulting much higher $\lambda$ for $\theta$-solvent. The variation of $\lambda$ with solvent quality is illustrated in inset (i) of Fig. 5(a). In $\theta$-solvent, despite the excluded volume effect being nullified, the strengthened elastic effect with increasing $c$ becomes strong enough to make the elasticity-induced disturbance as the dictating one resulting in an enhancement of the load capacity ratio ($\tilde{W}/\tilde{W}_{\text{ref}}$) up to ~ 1.13 times. Conversely, in a good solvent, the amplification in the degree of elasticity is much weaker, and therefore, despite the excluded volume effect being advantageous, here, load capacity ratio ($\tilde{W}/\tilde{W}_{\text{ref}}$) rises only up to ~ 1.03 times. So, one can conclude that, for achieving improved load bearing capacity, a dilute solution consisting $\theta$-solvent should be deployed over good solvent.

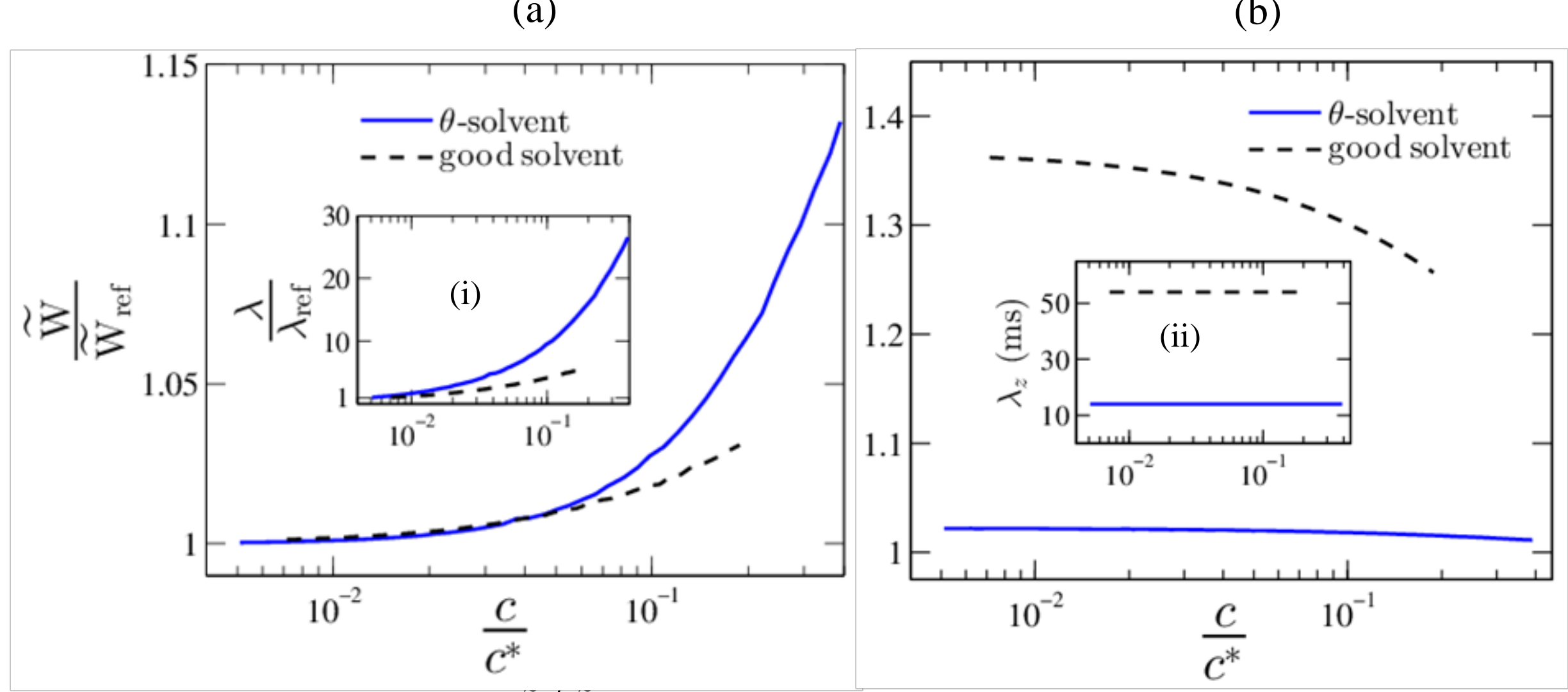

**Fig. 5.** The variation of $\tilde{W}/\tilde{W}_{\text{ref}}$ with varying solvent quality. (a) incorporates results using reported experimental data of fluid relaxation times, (b) predictions according to the Zimm's relaxation time. Here, Dioctyl Phthalate (DOP) and Tricresyl Phosphate (TCP) are chosen as examples of $\theta$-solvent and good solvents respectively while Polystyrene (PS) is chosen as the representative example of a neutral polymer ($M_w$ = 6.9 MDa). Inset (i) shows the variation of $\lambda$ with $c$ for different solvent quality while inset (ii) shows the corresponding $\lambda_z$ values.

Fig. 5(b) includes the predictions of load capacity ratio ($\tilde{W}/\tilde{W}_{\text{ref}}$) using Zimm's relaxation time ($\lambda_z$) which is independent of polymer concentration ($c$) for a dilute solution, while in reality, $\lambda$ depends on $c$ as $\lambda \propto c^{0.76}$ for $\theta$-solvent and $\lambda \propto c^{0.54}$ for good solvent respectively. [88] As a result, the use of $\lambda_z$ leads to grossly erroneous predictions of $\tilde{W}/\tilde{W}_{\text{ref}}$. As reported earlier, Zimm's theory may lead to discrepancies in finding $\lambda$ for good solvent which is also observed here in Fig. 5(b) by the opposite trend in the load capacity ratio compared to Fig. 5(a).

Now, the attention is given towards the polyelectrolyte solutions [88,91,92] which exhibit strikingly distinct behavior as compared to neutral polymer solutions. For neutral polymers, no interaction between the chains takes place below $c^*$, while for polyelectrolyte solutions, strong intermolecular interaction occurs at a much lower concentration than $c^*$ which results in swelling of the chain. With increasing $c$, strengthened interaction between polymer chains augments both fluid viscosity $(\mu)$ and relaxation time $(\lambda)$ where the growth of $\lambda$ is much faster as compared to $\mu$. As clear from inset (i) and (ii) of Fig. 6(a), the relative increment for $\mu$ is ~ 3 times and ~ 20 times for $\lambda$ respectively. Thus, the relative strength of the elastic effect as compared to the viscous effect increases. Here, non-uniformity in the flow field is already induced by virtue of axial modulations of slip and surface potentials, and the strengthened elastic stress results in stronger deformation of the polymer chains. Higher the value of $c$, higher is the elasticity-governed perturbation to the flow field thus rising the load capacity ratio ($\tilde{W}/\tilde{W}_{\text{ref}}$) with increasing $c$ (keeping electrolyte concentration $n_0$ constant) significantly up to ~ 4.14 times as shown in Fig. 6(a). Here, the rheological properties of dilute Hyaluronic acid (HA) in Phosphate buffered Saline (PBS) solution has been chosen as an example of polyelectrolyte solution [88]. Usually, in absence of electrolyte or in the low electrolyte concentration regime, electrostatic repulsion between the free charges of the polymer expands the polymer chain. With increasing electrolyte concentration, this interaction gets screened by the counterions which contract the polymer chain. This alteration in the molecular conformation significantly affects the rheological properties like fluid viscosity ($\mu$) and relaxation time ($\lambda$). Subsequently, increasing electrolyte concentration ($n_0$) from 50 mM to 138 mM (using NaCl as an electrolyte) results in ~ 2.69 times further increment of the load capacity ratio, evident from Fig. 6(a).

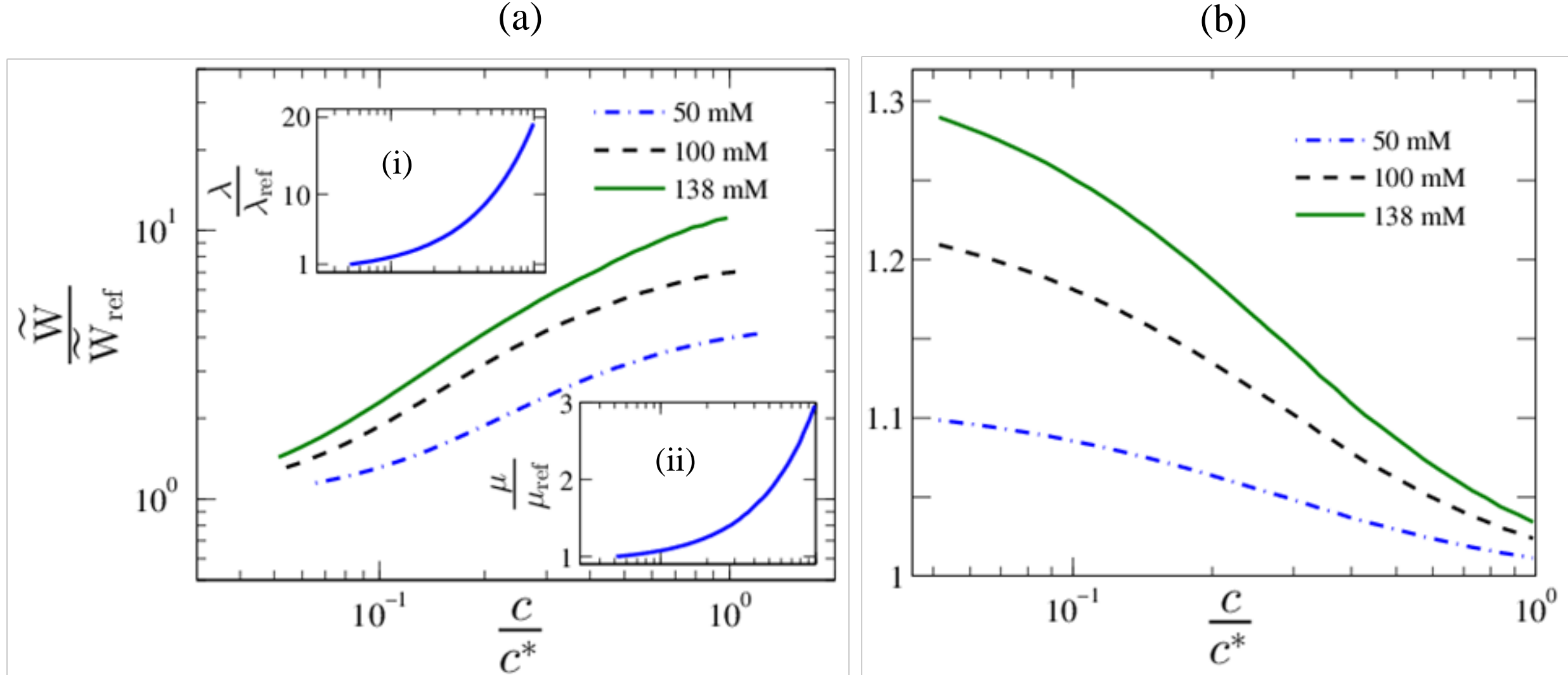

**Fig. 6**(a) - (b). The dependence of $\tilde{W}/\tilde{W}_{\text{ref}}$ with $c$ for different $n_0$ of electrolyte NaCl. It corresponds to the results using dilute Hyaluronic acid (HA) solution ($M_w$ = 1.6 MDa) in Phosphate buffered Saline (PBS) solution. (i) uses the expt. data for rheological parameters while (ii) uses predictions according to Zimm's theory. Inset (a) and (b) in (i) shows the variation of $\lambda$ and $\mu$ with increasing $c$ with $\lambda_{\text{Ref}}$ and $\mu_{\text{Ref}}$ being evaluated at $c/c^* = 0.066$.

Now, Fig. 6(b) shows the variation of same using theoretically estimated rheological parameters. As discussed earlier, the dependence of $\lambda$ in polyelectrolyte solution with $c$ is completely different as compared to neutral polymer (for which Zimm's theory is available). Since Zimm's definition of $\lambda$ fails to incorporate intermolecular interaction in polyelectrolyte solutions, it leads to erroneous predictions of load bearing capacity where an opposite trend of load capacity ratio ($\tilde{W}/\tilde{W}_{\mathrm{ref}}$) with $c$ is observed in Fig. 6 (b). Here, the important conclusion is that by selecting the proper concentration of electrolyte in polyelectrolyte solutions, an order of magnitude enhancement of load bearing capacity can be achieved in viscoelastic fluids as compared to the corresponding Newtonian fluid.

## 5. Conclusions

The primary objective of the present analysis is to delineate the possible sources of alteration in the hydrodynamics of a deformable microfluidic channel under the rheological premises of viscoelastic fluids, typically reminiscent of complex biological fluids. The complex interplay of elasto-mechanics, electrokinetics, hydrodynamics, and fluid rheology under a microfluidic environment has been analyzed. It is unveiled that by choosing a proper concentration, the molecular weight of a polymer, and the quality of the Newtonian solvent in case of neutral polymeric solutions and by modulating the electrolyte concentration in the case of polyelectrolyte solutions, it is practically possible to enhance the load bearing capacity of the microchannel up to one order of magnitude as compared to that of a Newtonian fluid. We understand that these outcomes may be beneficial in improved and optimal designing of lubricated systems involving biofluids thus bearing strong contemporary relevance.

**Declaration of Competing Interest**

There are no conflicts to declare.

**Acknowledgments**

One of the authors, SM, gratefully acknowledges the help of Dr. Jayabrata Dhar (Post-doctoral researcher, University of Luxembourg, 162 A, Avenue de la Faïencerie, L-1511 Luxembourg City, Luxembourg) regarding the problem formulation part.

## Appendix

### Section A: The coefficients involved in Eqs. (11) & Eq. (14)

The coefficients of Eqs. (11) and (14) are given below

$$\left[\chi_1 = a_{35}, \chi_2 = a_{37} + b_7, \chi_3 = a_{36} + b_8, \chi_4 = a_{38}\right]$$

$$\left[\begin{array}{l}
a_1 = -\frac{2}{3}\tilde{h}^3 - 2\tilde{l}_s\,\tilde{h}^2,\; a_2 = 2\beta\, a_{45}{}^2\, a_{52}\, a_{39}\left(\tilde{x}\right), a_4 = 2\,\beta\, a_{66}\, a_{39}\left(\tilde{x}\right) a_{41},\; a_5 = -2\, a_{66}\,\tilde{q}\, a_{40}\left(\tilde{x}\right) a_{41}\,\tilde{h}, \\
a_3 = 2\, a_{45}{}^2\, a_{52}\,\tilde{q}\, a_{40}\left(\tilde{x}\right)\left[a_{41}\, a_{50}\, a_{57}\left(\tilde{x}\right) - \tilde{h}\right],\; a_6 = 2\, a_{67}\, a_{39}\left(\tilde{x}\right) a_{42}\,\beta,\; a_7 = -2\, a_{67}\,\tilde{q}\, a_{40}\left(\tilde{x}\right) a_{42}\, a_{43}\left(\tilde{x}\right), \\
a_8 = 3\, a_{65}\,\beta,\; a_9 = a_8\, a_{43}\left(\tilde{x}\right)/\beta,\; a_{10} = -\left(3\, a_{65}\, a_{49}\, a_{42}/2\right), a_{14} = -a_{64}\,\tilde{h}^3 a_{47}, a_{11} = -\frac{3}{2} a_{65}\, a_{43}\left(\tilde{x}\right) a_{47}, \\
a_{12} = -\left(a_8\, a_{47}/2\right),\; a_{13} = -6\, a_{64}\, a_{49}\,\tilde{h}^2\, a_{42},\; a_{15} = -3\, a_{14},\; a_{16} = a_{10},\; a_{18} = a_{12}, a_{51} = \beta/\tilde{h}, a_{66} = a_{46}\, a_{54}, \\
a_{17} = a_{11},\; a_{19} = a_9,\; a_{20} = a_8,\; a_{34} = 4\, a_{64}\, a_{52}\, a_{39}\left(\tilde{x}\right) a_{44}\, a_{42}{}^3\,\beta,\; a_{67} = a_{52}\, a_{44}, a_{68} = a_{41}\, a_{42}, a_{50} = 1/\tilde{Q}, \\
a_{22} = -\left[3\, a_{63}\, a_{58}\,\tilde{q}\, a_{40}\left(\tilde{x}\right)/a_{55}\right]\tilde{h}\left(a_{48} + 2\, a_{68}\right),\; a_{21} = \left[3\, a_{63}\, a_{61}\left(\tilde{x}\right)/a_{55}{}^2\right]\left(a_{41}\, a_{48} + 2\, a_{42}\right) a_{57}\left(\tilde{x}\right), \\
a_{23} = \left[3\, a_{63}\, a_{58}\, a_{39}\left(\tilde{x}\right)\beta/a_{55}\right]\left(a_{48} + 2\, a_{68}\right),\; a_{28} = 6\, a_{63}\, a_{61}\left(\tilde{x}\right) a_{47}\, a_{50}\,\tilde{h}, a_{30} = -12\, a_{64}\,\tilde{l}_s\,\tilde{h}^2 a_{42}, \\
a_{25} = -\left[3\, a_{63}\, a_{61}\left(\tilde{x}\right)/a_{56}\right]\tilde{h}\; a_{62},\; a_{24} = \left[3\, a_{63}\, a_{61}\left(\tilde{x}\right)/a_{56}{}^2\right]\left(a_{41}\, a_{48} - 2\, a_{42}\right) a_{57}\left(\tilde{x}\right),\; a_{44} = \left(1 + a_{52}\right), \\
a_{29} = -6\, a_{63}\, a_{58}\, a_{39}\left(\tilde{x}\right)\, a_{47}\, a_{50}\,\beta,\; a_{26} = \left[3\, a_{63}\, a_{58}\, a_{39}\left(\tilde{x}\right)\beta/a_{56}\right] a_{62}, a_{64} = a_{63}\, a_{44}{}^2,\; a_{65} = a_{64}\, a_{49}{}^2, \\
a_{27} = -a_{28}\, a_{41}\, a_{50}\; a_{57}\left(\tilde{x}\right)/\tilde{h},\; a_{31} = -12\, a_{64}\, a_{66}\,\tilde{q}\, a_{40}\left(\tilde{x}\right) a_{68}\,\tilde{h}, a_{39}\left(\tilde{x}\right) = \cos\left(\tilde{q}\,\tilde{x} + \tilde{\theta}\right), a_{58} = a_{44}{}^2 a_{45}\, a_{53} \\
a_{32} = 12\, a_{64}\, a_{66}\, a_{39}\left(\tilde{x}\right) a_{68}\, a_{42}\,\beta,\; a_{33} = -4\, a_{64}\,\tilde{q}\, a_{40}\left(\tilde{x}\right) a_{67}\, a_{42}{}^3\, a_{43}\left(\tilde{x}\right), a_{40}\left(\tilde{x}\right) = \sin\left(\tilde{q}\,\tilde{x} + \tilde{\theta}\right), \\
a_{35} = a_1 + 2\left(a_9 + a_{10} + a_{11} - a_{14}\right) + a_{13} + a_{30}, a_{41} = \tanh\left(\tilde{Q}\,\tilde{h}\right), a_{42} = \tanh\left(\tilde{\kappa}\,\tilde{h}\right), a_{62} = a_{48} - 2\, a_{68}, \\
a_{38} = a_3 + a_5 + a_7 + a_{21} + a_{22} + a_{24} + a_{25} + a_{27} + a_{28} + a_{31} + a_{33}, a_{49} = 1/\tilde{\kappa},\; a_{43}\left(\tilde{x}\right) = \tilde{h} + \beta\,\tilde{p}_1\left(\tilde{x}\right), a_{56} = \tilde{Q} - 2\tilde{\kappa}, \\
a_{46} = \tilde{\kappa}\, a_{45},\; a_{47} = 1 - a_{42}{}^2, a_{48} = 1 + a_{42}{}^2,\; a_{53} = \tilde{\kappa}\, a_{52}, a_{54} = \tilde{\kappa}\,\tilde{l}_s{}^2,\; a_{55} = \tilde{Q} + 2\tilde{\kappa}, a_{52} = \tilde{\kappa}\,\tilde{l}_s, a_{45} = \tilde{\kappa}/\tilde{Q}, \\
a_{37} = 2\left(a_8 + a_{12}\right), a_{36} = a_2 + a_4 + a_6 + a_{23} + a_{26} + a_{29} + a_{32} + a_{34} + c_1 + c_2 + c_3 + c_4 + c_5 + c_6 + c_7 + c_8 + c_9, \\
a_{57}\left(\tilde{x}\right) = 1 - a_{51}\,\tilde{p}_1\left(\tilde{x}\right), a_{63} = s\,\mathrm{De}^2, a_{59} = a_{67}\, a_{46}\, a_{44},\; a_{60}\left(\tilde{x}\right) = a_{59}\, a_{39}\left(\tilde{x}\right),\; a_{61}\left(\tilde{x}\right) = a_{58}\,\tilde{q}\, a_{40}\left(\tilde{x}\right),
\end{array}\right]$$

$$\begin{bmatrix} b_1 = 2\beta \tilde{l}_s\, a_{43}\left(\tilde{x}\right),\; b_2 = -2\, a_{66}\, a_{39}\left(\tilde{x}\right) a_{41}\, \beta,\; b_3 = -2\, a_{67}\, a_{39}\left(\tilde{x}\right) a_{42}\, \beta,\; b_4 = 12\, a_{64}\, a_{43}\left(\tilde{x}\right) \beta\, {a_{42}}^2, \\ b_5 = -12\, a_{64}\, a_{66}\, a_{39}\left(\tilde{x}\right) a_{68}\, \beta\, a_{42},\; b_6 = -4\, a_{64}\, a_{67}\, a_{39}\left(\tilde{x}\right) {a_{42}}^3\, \beta,\; b_7 = b_1 + b_4,\, b_8 = b_2 + b_3 + b_5 + b_6, \end{bmatrix}$$

$$\begin{bmatrix} c_1 = -2\beta\, a_{52}\, {a_{45}}^2 a_{39}\left(\tilde{x}\right) + 2\, a_{51}\, a_{52}\, {a_{45}}^2\, a_{39}\left(\tilde{x}\right) a_{41}\, a_{50},\; c_2 = -2\beta\, a_{45}\, {a_{52}}^2 a_{39}\left(\tilde{x}\right) a_{41}, \\ c_3 = -12\, a_{64}\, \beta\, a_{45}\, {a_{52}}^2\, a_{39}\left(\tilde{x}\right)\, a_{42}\, a_{68},\; c_4 = \left[3 a_{63}\, a_{51}\; a_{60}\left(\tilde{x}\right) / {a_{55}}^2\right]\left(a_{41}\, a_{48} + 2\, a_{42}\right), \\ c_5 = -\left[3 a_{63}\, \beta\, a_{60}\left(\tilde{x}\right) / a_{55}\right]\left(a_{48} + 2\, a_{68}\right),\; c_6 = \left[3 a_{63}\, a_{51}\, a_{60}\left(\tilde{x}\right) / {a_{56}}^2\right]\left(a_{41}\, a_{48} - 2\, a_{42}\right), \\ c_7 = -\left[3 a_{63}\, \beta\, a_{60}\left(\tilde{x}\right) / a_{56}\right] a_{62},\; c_8 = -6\, a_{63}\, a_{51}\, a_{60}\left(\tilde{x}\right) {a_{50}}^2\, a_{41}\, a_{47},\; c_9 = 6\, a_{64}\, \beta\, a_{52}\, a_{46}\, a_{39}\left(\tilde{x}\right) a_{50}\, a_{47} \end{bmatrix}$$

## Section B: Theoretical background

It is a well-known fact that the addition of a small amount of polymer drastically alters the rheological behavior of a Newtonian solvent. This addition initiates several thermodynamic interactions which result in a coil-type of a configuration of the polymer chain. [79,80] Now, the configuration is dictated by the competition of two counteracting forces - one is the steric repulsion while the other one being the solvent-mediated attraction between the monomers. The temperature of the solvent at which these two forces balance each other is termed as $\theta$-temperature and the corresponding solvent is $\theta$-solvent. As the temperature is increased, steric repulsion predominates over solvent-mediated attraction and the solvent is termed a good solvent. In practice, one can determine the quality of a solvent (whether good solvent or $\theta$-solvent) by using a parameter $\nu$, known as fractal polymer dimension, ranging from 0.5 ($\theta$-solvent) to 0.6 (good solvent). Now, depending on the addition of polymer, significant alteration in the rheological behavior of the aqueous solution can be observed and the solution may undergo a transition from one regime to another. The threshold concentration above which the polymer coils start to interact with each other is known as cross-over concentration or overlap concentration $\left(c^*\right)$. Below $c^*$, the solution can be assumed to be dilute for which the relaxation time is determined using the well-known Zimm's relaxation time [80,93]

$$\lambda_z = \frac{1}{\zeta\left(3\nu\right)} \frac{\left[\mu\right] M_w \eta_s}{N_A\, k_B\, T} \tag{A1}$$

where $\left[\mu\right]$ is the intrinsic viscosity, $\left(M_w\right)$ molecular weight, $\eta_s$ solvent viscosity, $N_A$ Avogadro's number, $k_B$ Boltzmann constant, $T$ absolute temperature and $\zeta\left(3\nu\right)$ is a coefficient accounted for the solvent quality. As concentration is increased beyond $c^*$, a transition from dilute to semi-dilute untangled regime takes place because of interaction between the polymer chains. In this regime, rheological properties can be obtained using Rouse theory in which the scaling laws are given below[79]

$$\mu_{\text{sp,semi-dilute}} \propto c^{1/(3\nu-1)} \quad \text{and} \quad \lambda_{\text{semi-dilute}} \propto c^{(2-3\nu)/(3\nu-1)}$$

Here, $\mu_{sp}$ is specific viscosity defined by $\mu_{sp} = \frac{\mu}{\mu_s} - 1$ where $\mu_s$ is the solvent viscosity and $\mu$ is the solution viscosity. In this regime, polymer coils interact without forming entanglement up to concentration $c_e$ (i.e., entanglement concentration) beyond which another transition occurs from semi-dilute unentangled to semi-dilute entangled, for which the scaling law is as follows [79]

$$\mu_{sp,\text{entangled}} \propto c^{3/(3\nu-1)} \text{ and } \lambda_{\text{entangled}} \propto c^{3(1-\nu)/(3\nu-1)}$$

**Section C: Intrinsic viscosity of dilute polymer solution**

The intrinsic viscosity [80,90,93,94] of the dilute polymer solution is defined as $[\mu] = \lim_{c\to 0} \frac{\eta - \eta_s}{c\,\eta_s}$. Experimentally, one can obtain the intrinsic viscosity of polymer solution by simply preparing series of solutions with enhancing the degree of diluteness and then extrapolating the ratio of $\frac{\eta - \eta_s}{c\,\eta_s}$ for zero concentration of polymer. This also shows a strong dependence with polymer molecular weight $(M_w)$ which reads as $[\mu] = a\,M_w{}^b$ (this is also known as Mark-Houwink-Sakurada (MHS) equation) where *a* and *b* are empirical constants that are obtained based on experimental data. For example, in the case of aqueous PEO solution, $[\mu] = 0.072\,M_w{}^{3\nu-1}$ with $\nu = 0.55$. [80] Once, intrinsic viscosity is obtained, one can obtain the overlap concentration either according to Flory's theory[87] $c^* = 1/[\mu]$ or Graessley's classification [95] $c^* = 0.77/[\mu]$. For example, in the case of aqueous PEO solution, $[\mu] = 0.072\,M_w{}^{3\nu-1}$ with $\nu = 0.55$.

**Section D: Limitations of Zimm's relaxation time**

The employment of Zimm's definition of relaxation time tells us that for a dilute solution, this is independent of the polymer concentration. In this context, it is important to mention that proper rheological characterization of viscoelastic fluids has become extremely important lately owing to their larger applicability in bio-microfluidic applications because of having remarkable similarities of their constitutive behavior with biofluids. So, accurate measurement of relaxation time is necessary as far as the designing and functioning of these devices consisting of these fluids are concerned. Nevertheless, the value of $\lambda$ is often found to be of the order of few milliseconds which is hard to determine and often beyond the scope of conventional rheometers. To circumvent this difficulty, some alternative techniques have been efficiently developed recently, namely capillary filament thinning methods (CaBER) and microfluidic approaches of determination [80,96,97].